\documentclass[]{interact}

\usepackage{amssymb}
\usepackage{amsfonts}
\usepackage{amsmath}
\usepackage{graphicx}
\usepackage{epstopdf}

\begin{document}

\title{Stopping of very heavy ions in Mylar}

\author{
\name{R.~N.~Sagaidak\thanks{CONTACT Email: sagaidak@jinr.ru}}
\affil{Flerov Laboratory of Nuclear Reactions, Joint Institute for Nuclear Research,\\ J.-Curie str. 6, 141980 Dubna, Moscow region, Russia}
}

\maketitle

\begin{abstract}
Available experimental data on Mylar stopping powers ($SP$s) for heavy ions (HIs) at energies $0.04 \leqslant E/A \leqslant 15$ MeV/nucleon have been compared to different semi-empirical model calculations with the aim of their possible usage for the estimates of ranges for very heavy ions at $E/A < 0.5$ MeV/nucleon, which are of practical interest. Significant deviations from the calculated $SP$ values were found for fission fragments and lighter HIs at $E/A < 1$ MeV/nucleon. A new model parameterization for Mylar $SP$ has been proposed. Range estimates obtained with any $SP$ model calculation show a critical dependence of their mean values on the approximated electronic stopping powers and the nuclear (collisional) $SP$ component. The last plays a crucial role at the end of the range and could only be obtained by calculations. Practical applicability of the results of investigation for very heavy evaporation residues (products of complete fusion reactions induced by HIs) implies the use of a thick catcher foil corresponding to the largest ranges derived in the estimates or the range measurements for these products.
\end{abstract}


\section{Introduction}
\label{intro}

In practice, we often need ranges of heavy evaporation residues (ERs) that produce complete fusion reactions induced by heavy ions (HIs) to assess the effectiveness of commissioning setups designed for the separation and detection of these extremely heavy ions. A gas-filled solenoid GASSOL \cite{gassol} is an example of such a setup, which is scheduled to go into operation in the near future. The Mylar foil of a defined thickness is a good choice as a catcher for ERs collection in the focal plane of the solenoid. Typical energies for Rn to Th ERs produced in test reactions with medium-heavy ions (from Ar to Ni) are below 0.5 MeV/nucleon. SRIM calculations \cite{SRIM} show very narrow range distributions for mono-energetic Rn to Th ions of respective energies in Mylar. This implies the use of a foil of a definite thickness accompanied by some pre-stopper for unambiguous identification of the collected reaction products, such as $\alpha$-radioactive ERs in particular. Obviously, the real range straggling (width of the distribution) is larger due to the evaporation of light particles from the compound nucleus formed in the chosen reaction and ERs stopping inside a target used in an experiment \cite{Saga13}.

In the concept of effective charge used in many semi-empirical approaches, the stopping power ($SP$) of a given medium for a given HI can be expressed through the $SP$ for a reference ion, and effective charges of HI and a reference ion, $Z^{*}$, at the same velocity. Representative data for light ions (H or He) and their approximations are usually used as references \cite{SRIM}, and thus $SP$ for a given HI can be written as
\begin{equation}\label{effchar}
  SP_{\rm HI} = SP_{\rm H} (Z^{*}_{\rm HI} / z^{*}_{\rm H})^2.
\end{equation}
In application of the effective charge concept, the effective charge fraction $\gamma = Z^{*}_{\rm HI} / Z_{\rm HI}$ ($Z_{\rm HI}$ is the atomic number of HI) is determined using measured $SP$ data \cite{IAEAsp}, so that Eq.~(\ref{effchar}) can be rewritten as
\begin{equation}\label{gamma}
  \gamma = z^{*}_{\rm H} / Z_{\rm HI} (SP_{\rm HI} / SP_{\rm H})^{1/2}.
\end{equation}
Thus obtained ``experimental'' $\gamma$ values are often approximated by an appropriate exponential function of velocity, which can be further used for the $SP$ determination of the medium for any HI at any velocity.

The $SP$ dependence on energy can be integrated, meaning that $SP \equiv -dE/dx$ (i.e., energy losses), so that the range of HI at energy $E$ can be obtained as
\begin{equation}\label{range}
  R = \int_{0}^{E} dE/SP_{\rm HI}.
\end{equation}
It should be noted that any range data in Mylar are unavailable, so the only way to get some idea of these values is to use available $SP$ data \cite{IAEAsp} for their subsequent approximation and integration as described above.

This work was motivated by the need to have reliable estimates of the ranges in Mylar for the very heavy ions of ERs with $Z_{\rm HI} \geqslant 90$ at energies corresponding to their production in complete fusion reactions induced by medium-heavy ions ($E/A < 0.5$ MeV/nucleon). First, the available Mylar $SP$ data for HIs are compared to existing semi-empirical model calculations, and a new semi-empirical parameterization for the $SP_{\rm HI}$ is attempted to be proposed. The range distributions of heavy ERs in Mylar are finally estimated using integrated approximations for $SP$ data and taking into account the energy spread of ERs escaping the target of finite thickness.

\section{Comparing SP data with model calculations}
\label{comparison}

\subsection{Two-parameter $\gamma$-approximation}
\label{2parappr}

In \cite{Saga15}, we performed an attempt to parameterize available Mylar $SP$ data for HIs, followed by their extrapolation to the heaviest atoms ($Z_{\rm HI} > 110$). A similar approach was adopted in the present work, but with the data \cite{Yu07,Dib15,Materna21}, which were not included in the previous approximation. Some minor corrections relating to $z^{*}_{H}$ and $SP_{H}$ were also made with respect to the previous work \cite{Saga15}, in which Ziegler's approximation \cite{Ziegler80} for $z^{*}_{H}$ and ``the normalization factor'' for $SP_{H}$ according to \cite{Barbui10par} were used.

\begin{figure}[h]  
\centering
\includegraphics[width=0.85\textwidth]{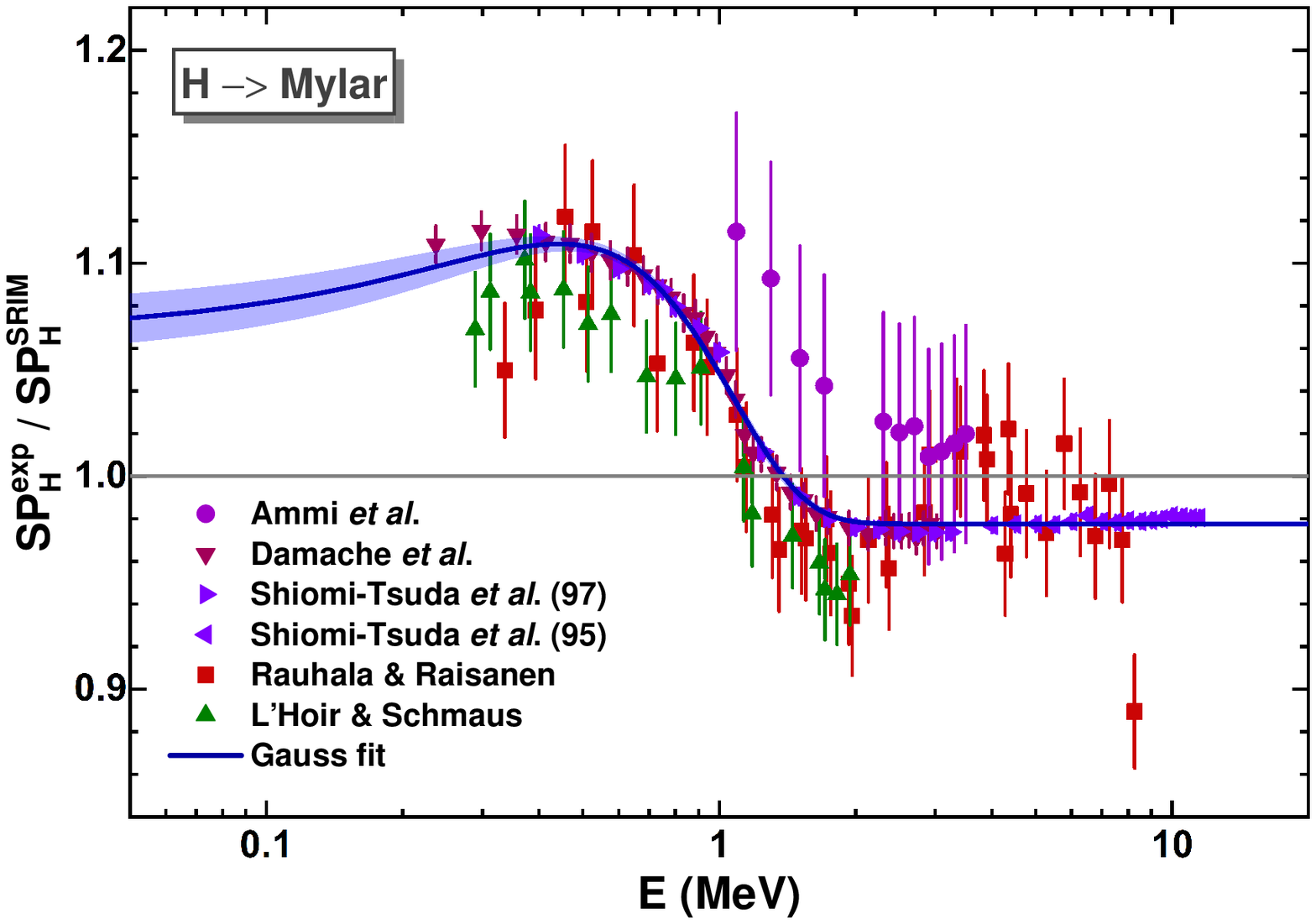}
\vspace{-3.5mm}
\caption{The ratios of the measured (see \cite{IAEAsp} for the original references designated by the first author's names) and SRIM calculated \cite{SRIM} $SP_{H}$ values as a function of energy (symbols). The best fit to the data is shown by a solid line (see the text for details).}\label{SPHratio}
\end{figure}

In the present work, the approximation to the theoretical curve, corresponding to the energy stability condition \cite{Yarlagadda78}, and good agreement with experimental data led to the relation:
\begin{equation}\label{ZeffH}
  z^{*}_{\rm H} = 1 - \exp[0.12751(V/v_{0})^{1/2} - 0.54485(V/v_{0}) - 0.46526(V/v_{0})^2],
\end{equation}
where $(V/v_{0})$ is the ratio of the hydrogen and Bohr velocities. Fig.~\ref{SPHratio} shows the ratios of the measured \cite{IAEAsp} and SRIM calculated \cite{SRIM} $SP_{\rm H}$ values.
The best fit to the $SP^{exp}_{\rm H} / SP^{\rm SRIM}_{\rm H}$ data was obtained with the function:
\begin{equation}\label{spHfit}
  SP^{exp}_{\rm H} / SP^{\rm SRIM}_{\rm H} = 0.9775 + 0.1316 \exp\{-0.5 [(E - 0.4428)/0.5008]^{2}\},
\end{equation}
where the hydrogen energy is in MeV. Note that, in contrast to the polynomial fit proposed in \cite{Barbui10par}, Eq.~(\ref{spHfit}) better describes the ratios outside the data range. This is confirmed by the 95\% confidence interval shown in Fig.~\ref{SPHratio} in the shadowed area.

\begin{figure}[h]  
\centering
\includegraphics[width=0.85\textwidth]{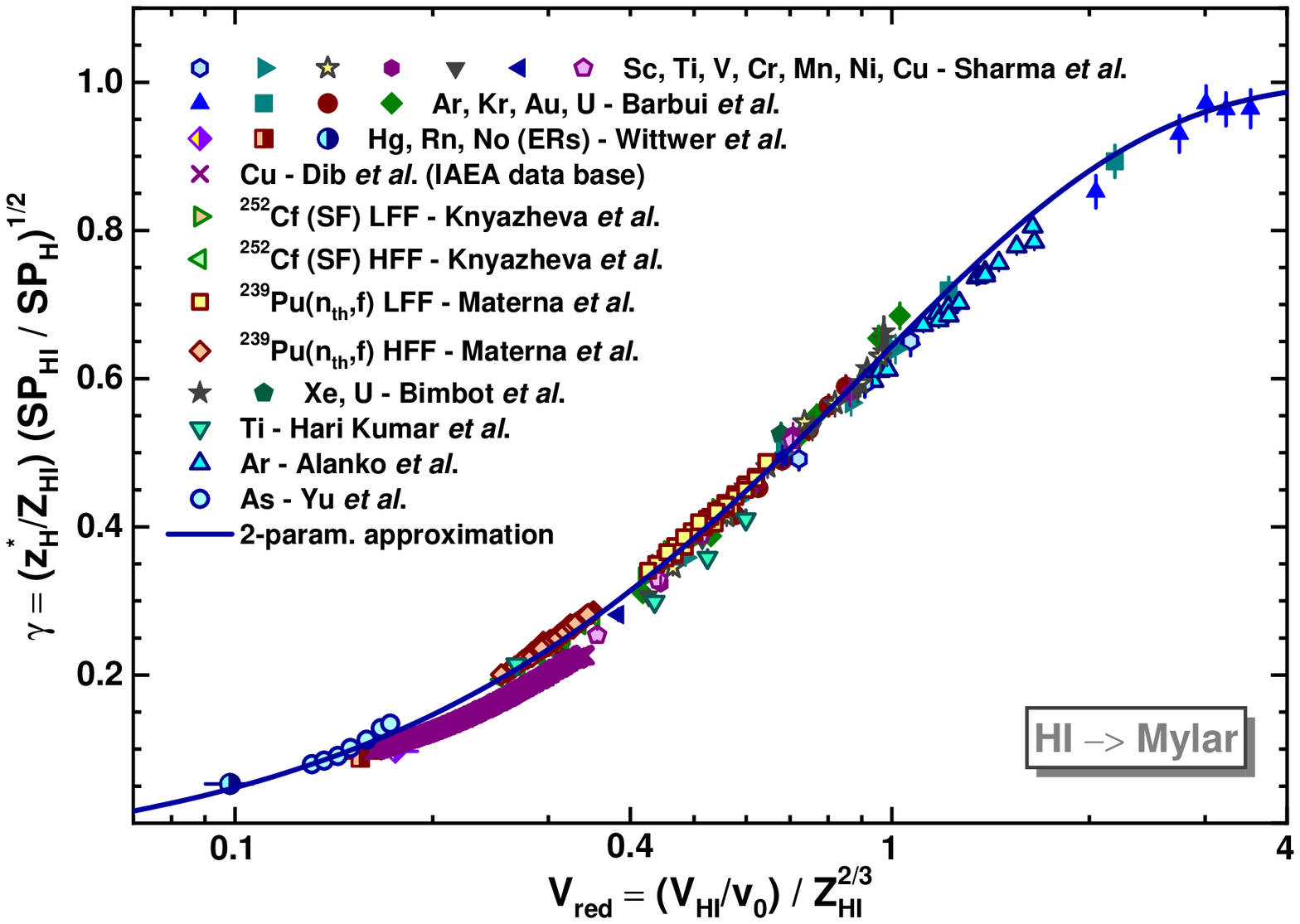}
\vspace{-2.0mm}
\caption{$\gamma$ values derived from the available data for $SP_{\rm HI}$ (see \cite{IAEAsp,Wittwer10} for the original references designated by the first authors' names) are shown as a function of the HI reduced velocity (symbols). The best fit to the data using Eq.~(\ref{apgamVred}) is shown by a solid line.}\label{gamVred}
\end{figure}

Using Eq.~(\ref{ZeffH}) for the effective charge of hydrogen and Eq.~(\ref{spHfit}) for the correction to the SRIM stopping powers, $\gamma$ values derived from the available data for $SP_{\rm HI}$ \cite{IAEAsp,Wittwer10} can be presented as a function of the HI reduced velocity $V_{red} = (V_{\rm HI}/v_{0}) / Z_{\rm HI}^{2/3} \simeq 6.35(E/A)^{1/2} / Z_{\rm HI}^{2/3}$. They are shown in Fig.~\ref{gamVred}. Here and below, stopping powers for heavy ions with $Z_{\rm HI} \geqslant 18$ are considered. The dependence $\gamma = f(V_{red})$ thus obtained was approximated by the two-parameter function, i.e., a similar one as used in \cite{Saga15}:
\begin{equation}\label{apgamVred}
  \gamma_{2p} = 1 - 1.0620 \exp(-1.0914 V_{red}).
\end{equation}

\begin{figure}[h]  
\centering
\includegraphics[width=0.85\textwidth]{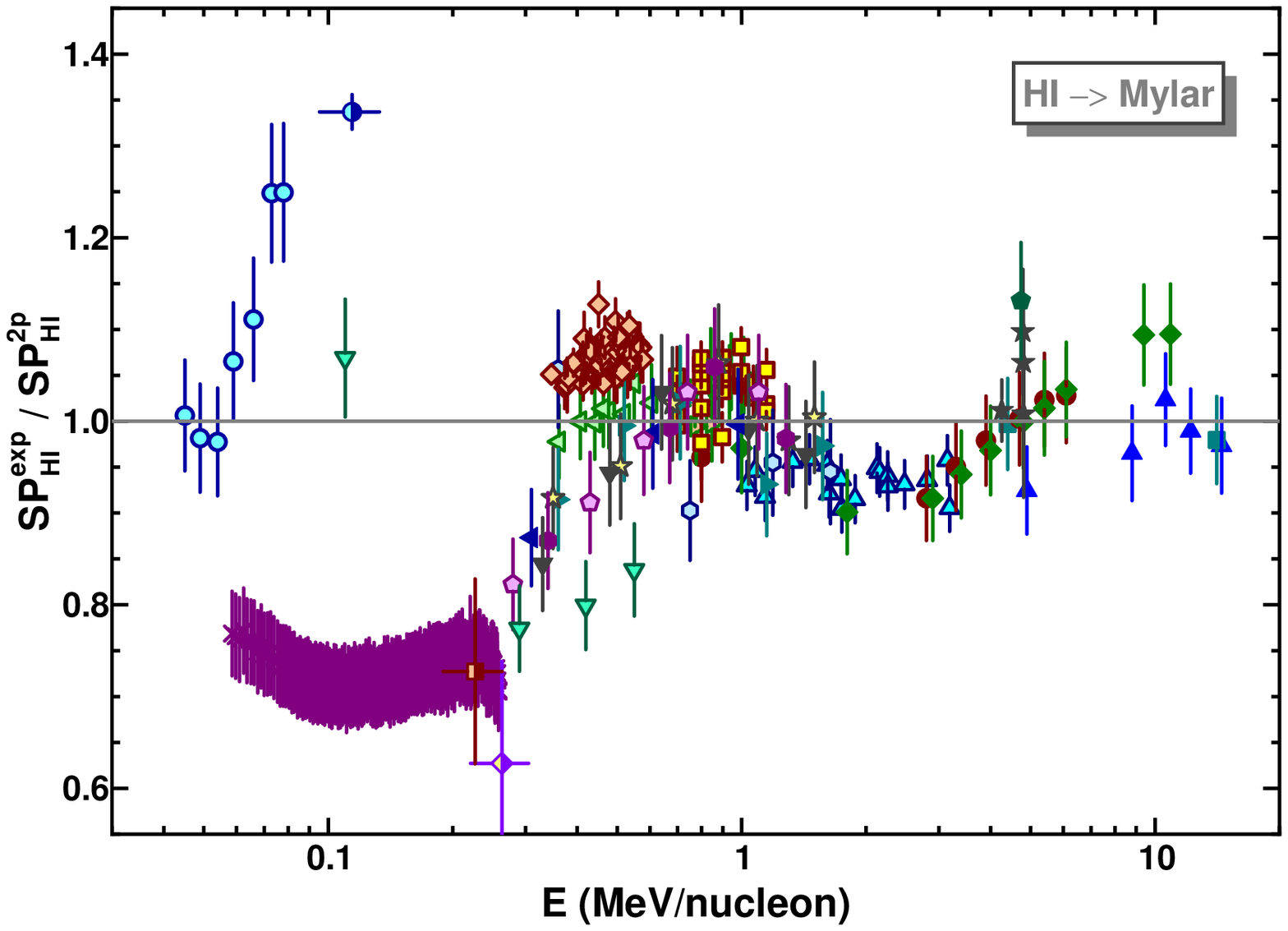}
\vspace{-2.0mm}
\caption{Comparing empirical $SP_{\rm HI}^{2p}$ values corresponding to the 2-parameter approximation using Eqs.~(\ref{ZeffH})--(\ref{apgamVred}) with the available data for $SP_{\rm HI}^{exp}$ \cite{IAEAsp,Wittwer10} is shown as a function of the HI energy using the same symbols as in Fig.~\ref{gamVred} for the data points.}\label{SP2pcomp}
\end{figure}

With the 2-parameter $\gamma$-approximation,  Eqs.~(\ref{ZeffH}) and (\ref{spHfit}) for $z^{*}_{\rm H}$ and correction of the SRIM stopping power for hydrogen, respectively, one may estimate empirical $SP_{\rm HI}^{2p}$ values and compare them with the experimental data \cite{IAEAsp,Wittwer10}. The result of such a comparison is shown in Fig.~\ref{SP2pcomp} as the ratio of both values. We see that deviations of the approximated data from experimental ones achieve more than 30\% at $E/A < 0.3$ MeV/nucleon for HIs of different $Z$. Such uncertainties in the prediction of $SP_{\rm HI}$ do not make it possible to obtain reliable ranges for very heavy ions.

\subsection{Five-parameter $\gamma$-approximation \cite{Barbui10par}}
\label{5parappr}

\begin{figure}[h]  
\centering
\includegraphics[width=0.85\textwidth]{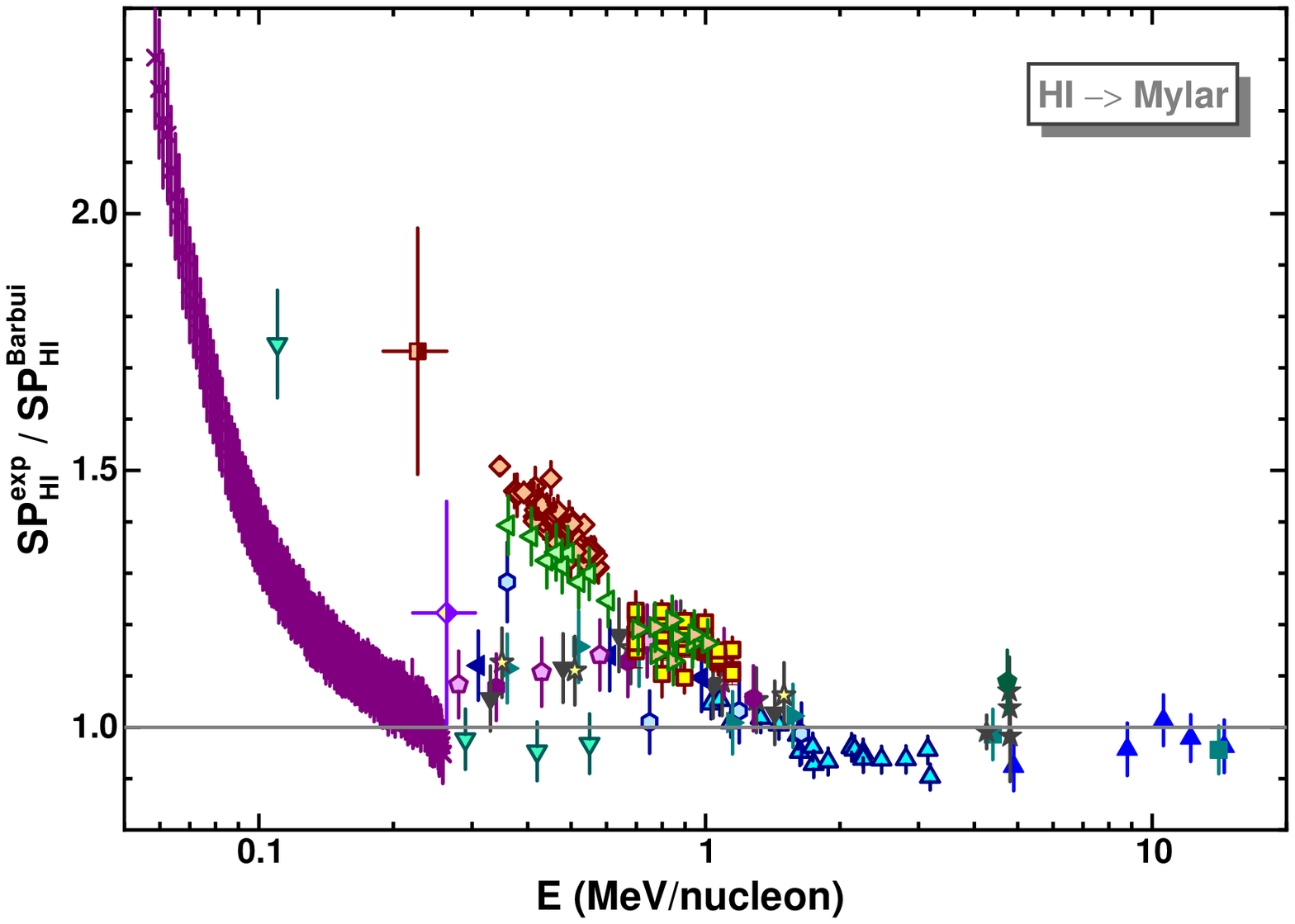}
\vspace{-2.0mm}
\caption{The same as in Fig.~\ref{SP2pcomp}, but for the approximation proposed by Barbui et al. \cite{Barbui10par}. The data \cite{Yu07,Wittwer10} exceed the upper boundary of the $SP_{\rm HI}^{exp} / SP_{\rm HI}^{\rm Barbui}$ range in the figure.}\label{Barbcomp}
\end{figure}

The earlier $\gamma$-approximation cited above \cite{Barbui10par} followed the experimental data obtained by this group for Ar, Kr, Au, and U ions \cite{Barbui10} (see Fig.~\ref{gamVred}). They used $z^{*}_{\rm H} = 1$ and a polynomial approximation for the correction of the SRIM stopping power for hydrogen (see \cite{Barbui10par} for details). The 5-parameter $\gamma$-approximation was proposed, which is written for Mylar as:
\begin{equation}\label{Barbpar}
  \gamma = 1 - [1.45 - 0.07\ln(Z_{\rm HI})] \exp[-6.0 (E/A)^{0.5089} / Z_{\rm HI}^{0.62}].
\end{equation}
In Fig.~\ref{Barbcomp}, the experimental data \cite{IAEAsp,Wittwer10} are compared to the approximated ones as the ratio of both values.

In the framework of this approximation, a spread of $\pm$5\% was obtained \cite{Barbui10par} for their data \cite{Barbui10}, together with some others. At the same time, this approximation significantly underestimates stopping powers for heavy fission fragments (designated as HFF in Fig.~\ref{gamVred}) and for medium-heavy ions Ti \cite{HariKumar94}, Cu \cite{Dib15}, and As \cite{Yu07} at energies below 0.12 MeV/nucleon. The last data \cite{Yu07} are outside the $SP_{\rm HI}^{exp} / SP_{\rm HI}^{\rm Barbui}$ figure range.

\subsection{Semi-empirical approximation based on Bohr's theory \cite{Knyazheva06}}
\label{Bohrappr}

In \cite{Knyazheva06}, the stopping power of fission fragments (FF) from $^{252}$Cf(sf) has been measured in thin foils, including Mylar, as a function of FF mass and energy. The measured $SP$ data for fission fragments cover masses within the range of $A$ = 101--148. The results are compared with semi-empirical predictions of the SRIM code, LSS theory, and theoretical calculations by Sigmund (PASS code). The best description of the data was achieved with a semi-empirical formula based on classical Bohr theory \cite{Bohr48}. The formula was also successfully tested on the available data for Ar, Ca, Kr, Xe, and Au ions at energies $E/A = 0.2-6$ MeV/nucleon slowing down in C, Ni, and Au.

According to the theory, the expression for electronic stopping power $SP_{e}$ for Mylar can be rewritten in the units of MeV/(mg/cm$^{2})$:
\begin{equation}\label{Knyazheva}
                SP_{e} = 0.1432 (Z_{m}/A_{m}) (Z_{\rm HI}^{*})^2 L(\xi) / (E/A),
\end{equation}
with the use of
\begin{eqnarray}
\nonumber  Z_{\rm HI}^{*} &=& Z_{\rm HI} \{1.0 - exp[-6.35 * (E/A)^{1/2} / Z_{\rm HI}^{0.57}]\}, \\
\nonumber                    L(\xi) &=& 1.1408 - 0.4565 * ln(\xi) + 0.04039 * [ln(\xi)]^2 + ln(\xi), \\
\nonumber                         \xi &=& 782.58 (E/A)^{3/2} / (Z_{\rm HI} I_{0} Z_{m}),
\end{eqnarray}
where $Z_{m} = 4.55$, $A_{m} = 9.09$ for Mylar, and $I_{0}\approx 10$ eV were used \cite{Knyazheva06}.
\begin{figure}[h]  
\centering
\includegraphics[width=0.85\textwidth]{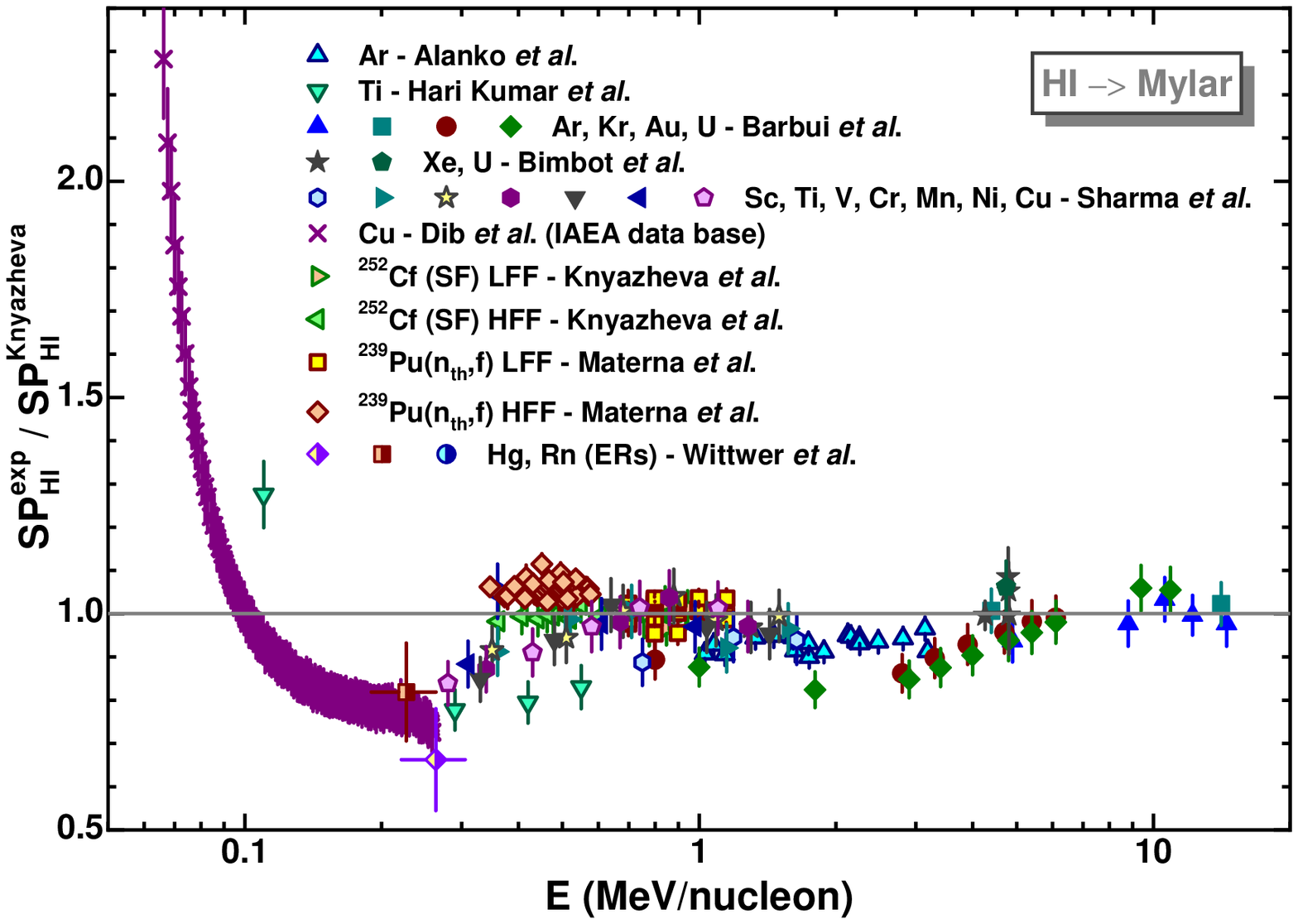}
\vspace{-2.0mm}
\caption{The same as in Figs.~\ref{SP2pcomp} and \ref{Barbcomp}, but for the semi-empirical Bohr approximation proposed in \cite{Knyazheva06}. The data \cite{Yu07,Wittwer10} exceed the upper boundary of the $SP_{\rm HI}^{exp} / SP_{\rm HI}^{\rm Knyazheva}$ range in the figure.}\label{Knyazhevacomp}
\end{figure}

Fig.~\ref{Knyazhevacomp} shows a comparison of the experimental data  \cite{IAEAsp,Wittwer10} with calculations according to Eq.~(\ref{Knyazheva}). As one can see, this approximation significantly underestimates stopping powers for the medium heavy ions Cu \cite{Dib15} and As \cite{Yu07} at energies below 0.08 MeV/nucleon, whereas, as might be expected, the fission fragment data \cite{Materna21,Knyazheva06} are reproduced quite well. The data \cite{Yu07} are outside the $SP_{\rm HI}^{exp} / SP_{\rm HI}^{\rm Knyazheva}$ figure range.

\subsection{SRIM approximation \cite{SRIM}}
\label{SRIMappr}

\begin{figure}[h]  
\centering
\includegraphics[width=0.85\textwidth]{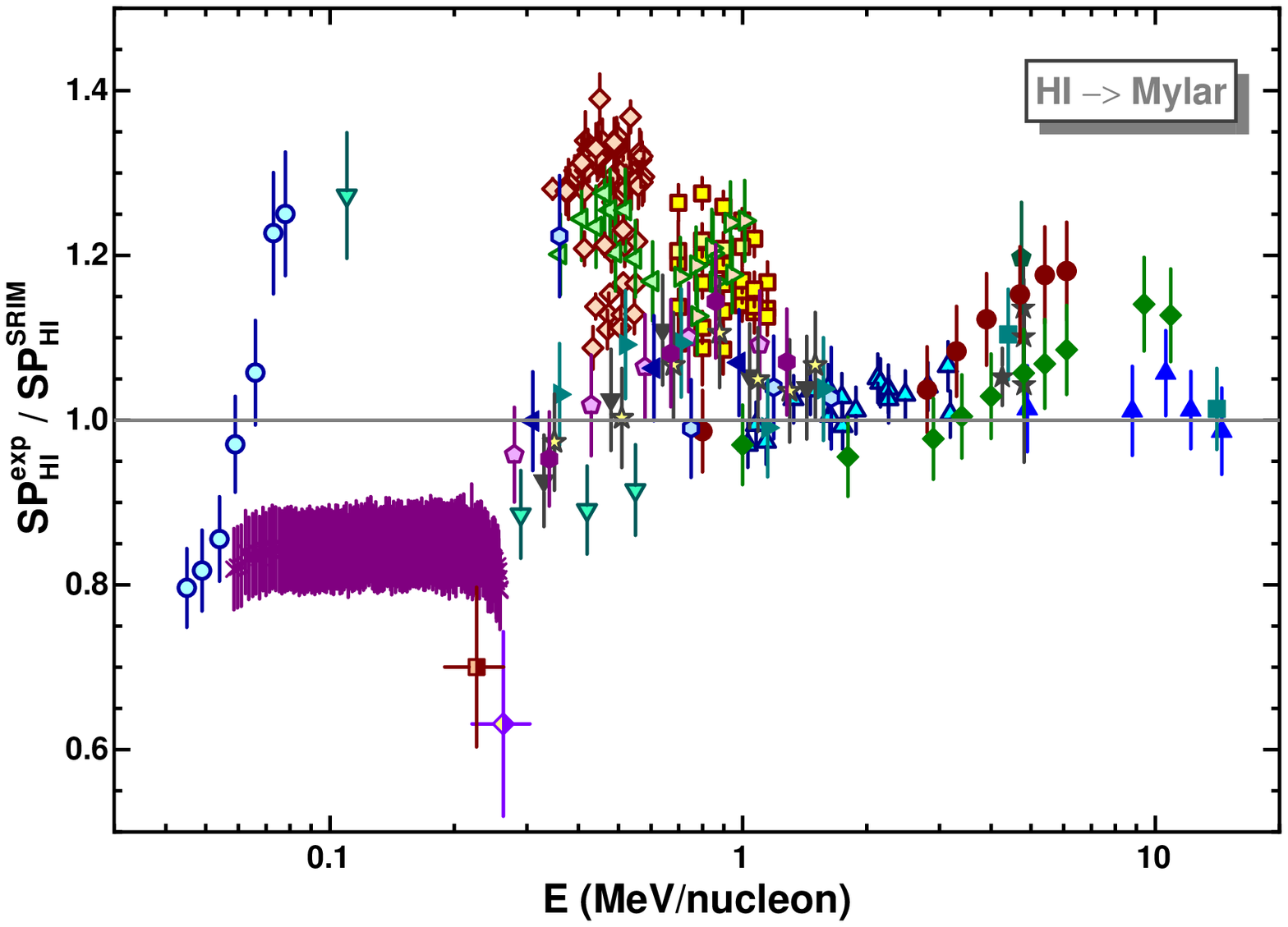}
\vspace{-2.0mm}
\caption{The same as in Figs.~\ref{SP2pcomp}, \ref{Barbcomp}, and \ref{Knyazhevacomp}, but for the SRIM calculations \cite{SRIM}.}\label{SRIMcomp}
\end{figure}

The well-known SRIM code \cite{SRIM} is widely applied to the estimation of stopping powers and ranges for HIs in different media. In Fig.~\ref{SRIMcomp}, a comparison of the experimental data  \cite{IAEAsp,Wittwer10} with SRIM calculations is shown. In contrast to previous approaches, the nuclear stopping power $SP_{n}$, calculated analytically, was taken into account in such a comparison. It implies that stopping powers for HIs are derived as follows:
\begin{equation}\label{SPtotal}
  SP_{\rm HI} = SP_{e} + SP_{n}.
\end{equation}
As we can see, SRIM on average reproduces experimental data, although the stopping power data for some groups of heavy fission fragments \cite{Materna21,Knyazheva06} are underestimated by a factor of 1.2--1.4. At the same time, SRIM stopping  powers for the low energy Cu ions \cite{Dib15} and very heavy Hg and Rn ERs \cite{Wittwer10} are overestimated by about the same factor.

\subsection{Empirical Ziegler's approximation \cite{Ziegler80}}
\label{Zieglerappr}

The analysis of the data presented in Fig.~\ref{SRIMcomp} suggests that SRIM reproduces differently in $SP$ for fission fragments with varying $Z$. Such an effect might be the result of a sharp dependency of $SP_{e}$ on $Z_{\rm HI}$ as it follows from low-energy SRIM calculations for HI stopping in carbon \cite{Saga23}. It was of interest to compare the experimental data with the earlier empirical approximation of Ziegler \cite{Ziegler80}, where a smooth dependency for effective charges is written as
\begin{equation}\label{Ziegler1}
  Z^{*}_{\rm HI} / z^{*}_{\rm H} = 1 - [\exp(-B)] [ 1.034 - 1.777\exp(-0.08114 Z_{\rm HI})],
\end{equation}
where
\begin{equation}\label{Ziegler2}
 B = 0.886 V_{red} + 0.0378 \sin(0.443 \pi V_{red}).
\end{equation}
For hydrogen, the corrected values of $SP_{\rm H}$ according to \cite{Barbui10par} were used. Using Eqs.~(\ref{effchar}), (\ref{Ziegler1}), and (\ref{Ziegler2}) for a definite HI velocity, electronic stopping powers $SP_{e}$ for HIs can be derived.

\begin{figure}[h]  
\centering
\includegraphics[width=0.85\textwidth]{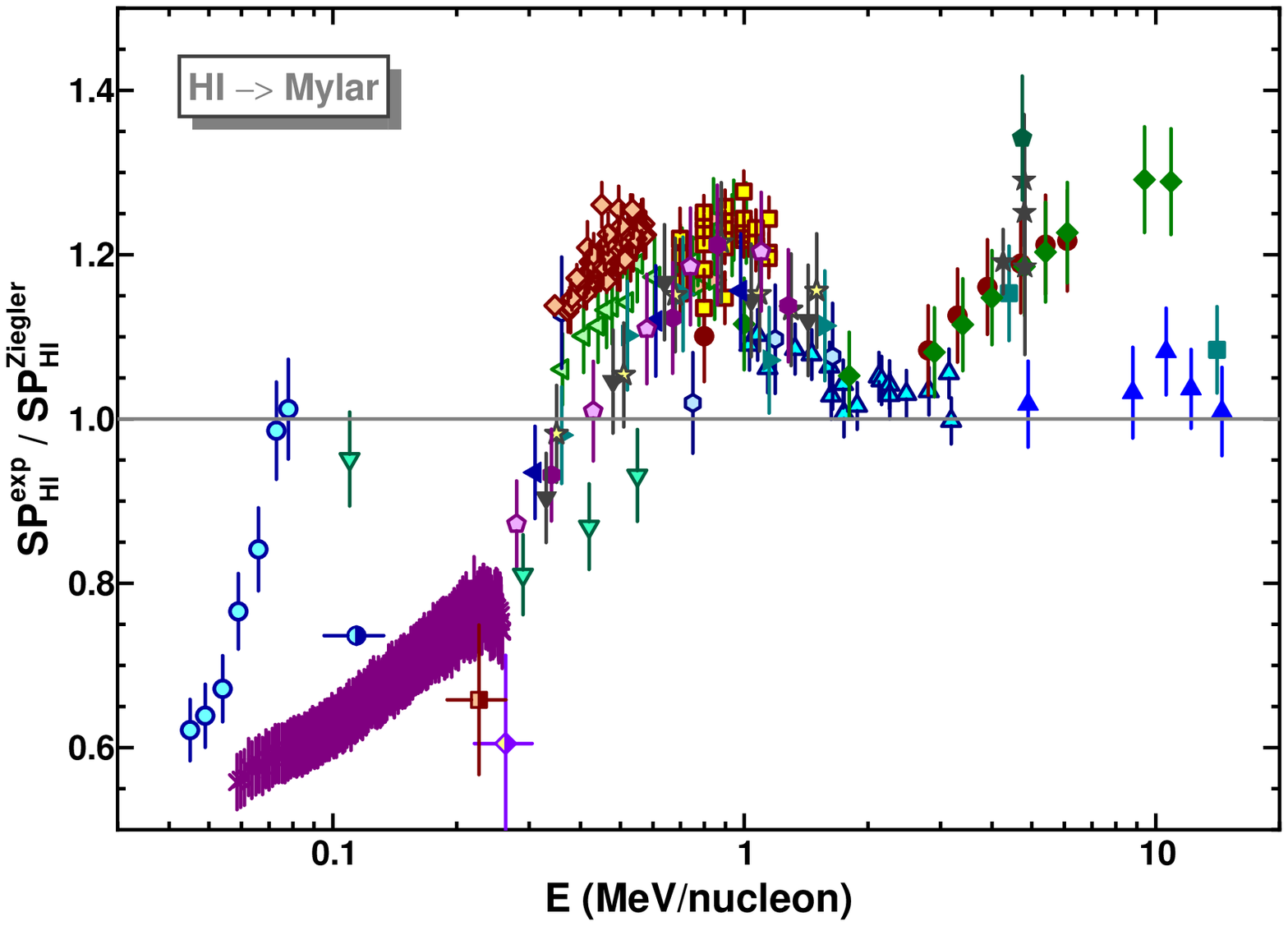}
\vspace{-2.0mm}
\caption{The same as in Figs.~\ref{SP2pcomp}, \ref{Barbcomp}, \ref{Knyazhevacomp}, and \ref{SRIMcomp}, but for Ziegler's approximations \cite{Ziegler80}.}\label{Zieglercomp}
\end{figure}

In this approximation, nuclear stopping power was taken into account. The theoretical value according to \cite{Wilson77} was used, which has the reduced stopping form:
\begin{equation}\label{WilsonSn}
  S_{n} = 0.5 \ln(1 + \varepsilon)/(\varepsilon + 0.10718 \varepsilon^{0.37544}),
\end{equation}
with the reduced energy defined as:
\begin{equation}\label{epsilon}
  \varepsilon = 32.53 A_{m} E / [Z_{\rm HI} Z_{m} (A_{\rm HI} + A_{m}) (Z_{\rm HI}^{2/3} + Z_{m}^{2/3})^{1/2}],
\end{equation}
where the HI energy, $E$, is in keV. Nuclear stopping powers $SP_{n}$ in units of MeV/(mg/cm$^{2}$) can be written as:
\begin{equation}\label{nuclearSP}
  SP_{n} = 0.5834 Z_{\rm HI} Z_{m} S_{n} / [(A_{\rm HI} + A_{m}) (Z_{\rm HI}^{2/3} + Z_{m}^{2/3})^{1/2}].
\end{equation}
Finally, stopping powers for HIs were derived from Eq.~(\ref{SPtotal}). The results of the comparison of thus calculated $SP_{\rm HI}^{\rm Ziegler}$ with experimental $SP_{\rm HI}^{exp}$ values are shown in Fig.~\ref{Zieglercomp}.

Note that the nuclear stopping calculated with Eqs.~(\ref{WilsonSn}) -- (\ref{nuclearSP}) is different from the one resulting from the SRIM application, as will be seen below. Often, the values of the nuclear-stopping additive are within the errors of the experimental $SP$ data. This additive becomes significant for very heavy ions at low energies.  For example, for $^{254}$No \cite{Wittwer10}, it makes up about 18\% of the $SP_{e}$ value, whereas for $^{75}$As at the lowest energy \cite{Yu07}, it makes up about 12\% of the $SP_{e}$ value, as it follows from the calculations \cite{Ziegler80}.  It may be recalled that the experimental $SP$ data \cite{IAEAsp,Wittwer10} correspond to the total stopping powers, whereas approximations may consider or not the nuclear stopping, as in the cases \cite{SRIM, Ziegler80} or \cite{Saga15,Barbui10par,Knyazheva06}, respectively. The last cases imply that nuclear stopping is reflected in low-energy data.

\subsection{New four-parameter $\gamma$-approximation}
\label{4parappr}

\begin{figure}[h]  
\centering
\includegraphics[width=0.85\textwidth]{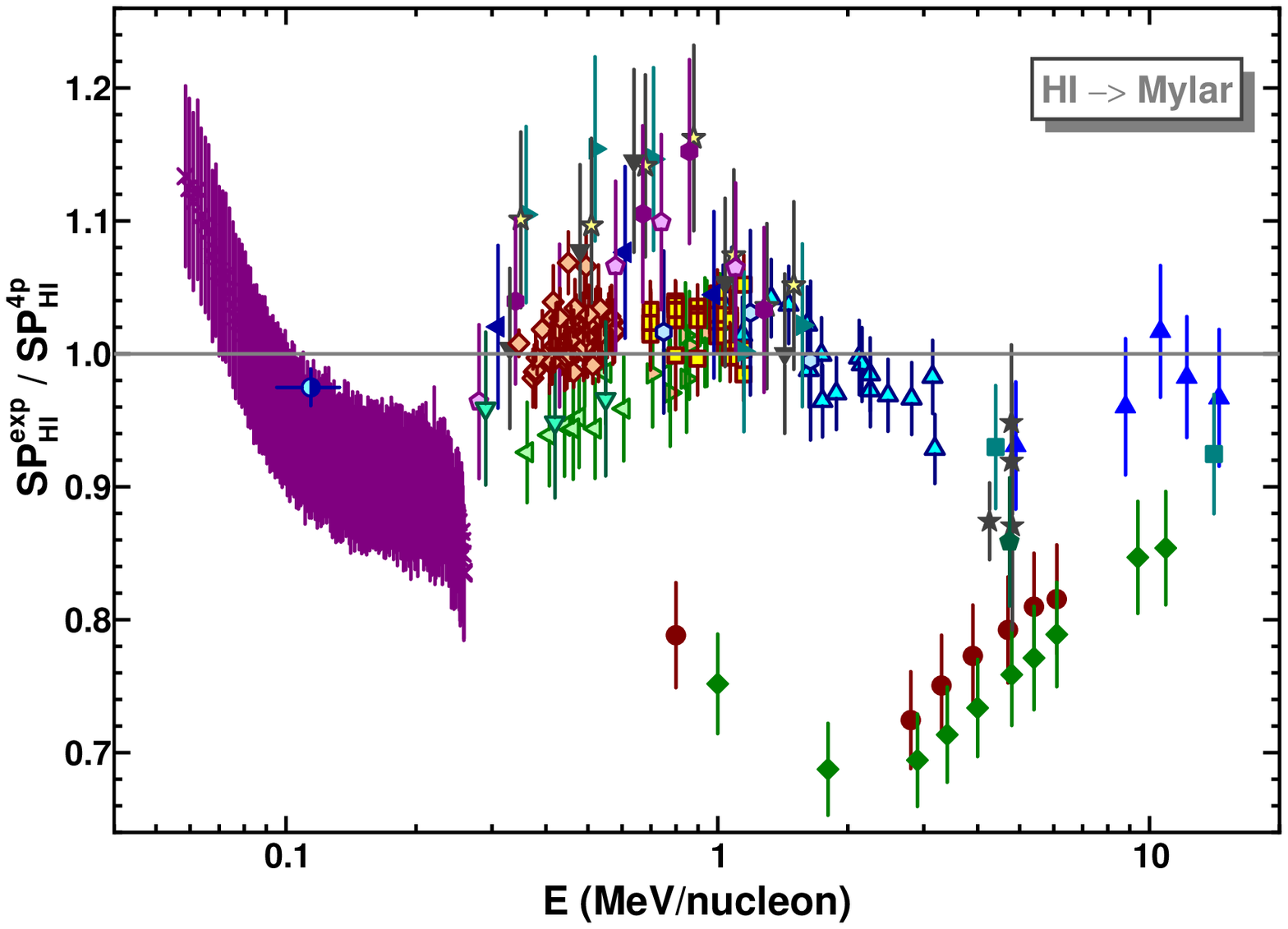}
\vspace{-2.0mm}
\caption{The same as in Figs.~\ref{SP2pcomp}, \ref{Barbcomp}, \ref{Knyazhevacomp}, \ref{SRIMcomp}, and \ref{Zieglercomp} but for the derived 4p-approximation.}\label{4pcomp}
\end{figure}

In light of poor data reproduction, as shown above, the data for $\gamma$-values obtained in Section~\ref{2parappr} were fitted with the 4-parameter exponential function expressed as
\begin{equation}\label{4pargamap}
  \gamma_{4p} = 1 - p_{0} \exp[p_{1} (E/A)^{p_{2}} / Z_{\rm HI}^{p_{3}}],
\end{equation}
where $p_{i}$ ($i$ = 0...3) are fitting parameters and $E/A$ is energy in MeV/nucleon. The following results, corresponding to the best fit were obtained: $p_{0} = 1.0583\pm0.0027, p_{1} = 4.309\pm0.102, p_{2} = 0.5475\pm0.0052$, and $p_{3} = 0.5349\pm0.0069$. Using this $\gamma_{4p}$-approximation and Eqs.~(\ref{ZeffH}) and (\ref{spHfit}) for $z^{*}_{\rm H}$ and the correcting function for $SP_{\rm H}^{\rm SRIM}$, respectively, one can compare the obtained $SP_{\rm HI}^{4p}$ values with the experimental $SP_{\rm HI}^{exp}$ data \cite{IAEAsp,Wittwer10}. The results of such a comparison are shown in Fig.~\ref{4pcomp}.

As can be seen from the figure, the proposed approximation reproduces almost all the data within a range of $\pm$20\%. The exceptions are the As data \cite{Yu07}, Au and U data at energies $1 \lesssim E/A \lesssim 6$ MeV/nucleon \cite{Barbui10}, and the data for Hg and Rn ERs \cite{Wittwer10}, for which the deviations exceed 20\%.

\subsection{Stopping powers for heavy ERs}
\label{SPforERs}

\begin{figure}[h]  
\centering
\includegraphics[width=0.85\textwidth]{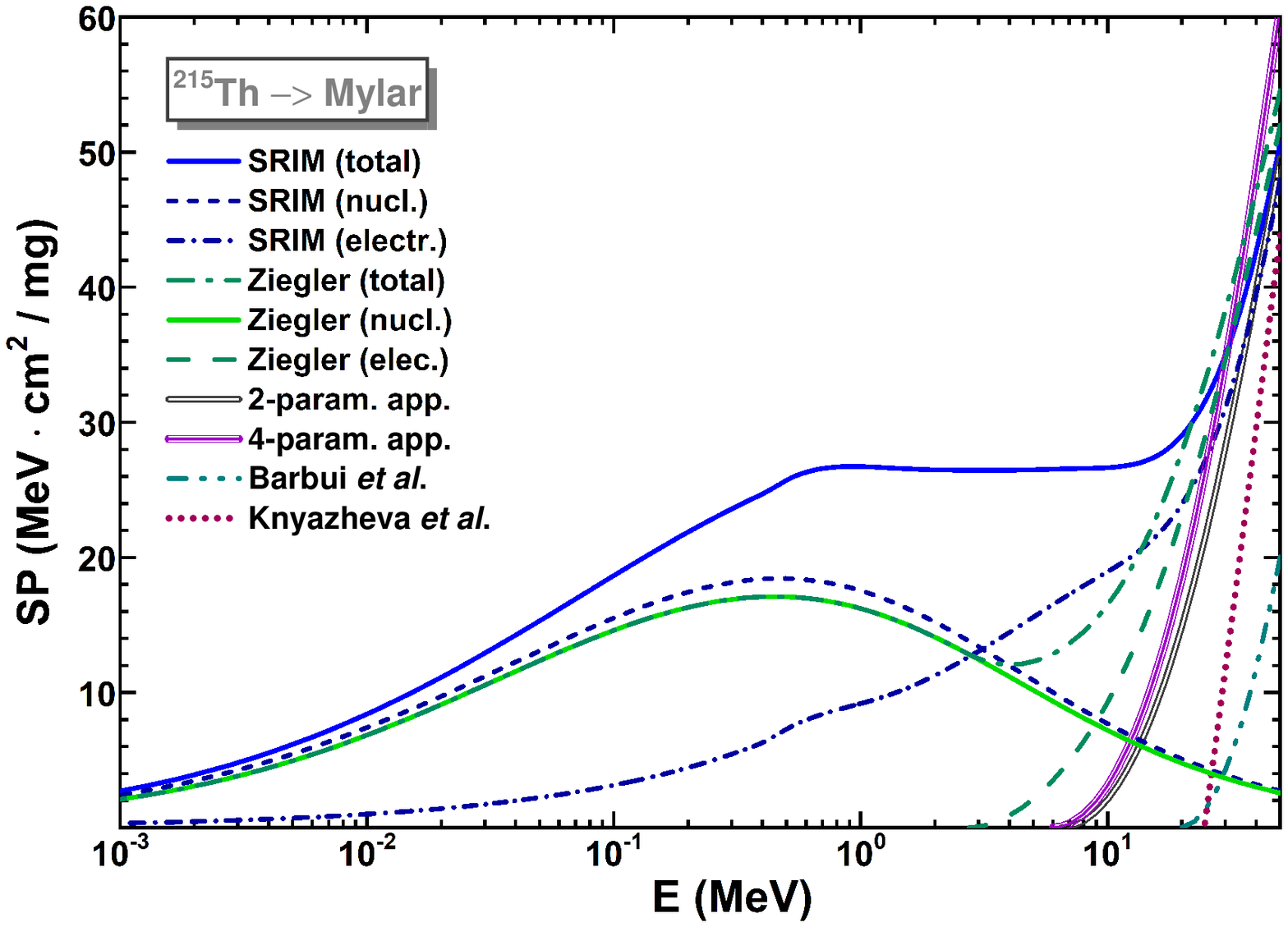}
\vspace{-2.0mm}
\caption{The Mylar stopping powers for $^{215}$Th, as obtained in the approximations considered in Sections~\ref{2parappr} -- \ref{4parappr} (different lines identified in the figure).}\label{215ThSP}
\end{figure}

As mentioned above, typical energies for Rn to Th ERs produced in test reactions with medium-heavy ions (from Ar to Ni) are below 0.5 MeV/nucleon. In Fig.~\ref{215ThSP}, the Mylar stopping powers calculated in the frameworks of the approximations considered in Sections~\ref{2parappr} -- \ref{4parappr} are shown for $^{215}$Th that can be produced in fusion reactions using respective targets.

One can see from the figure that 2p-, 4p-, and Ziegler's $SP_{e}$ \cite{Ziegler80} approximations show close values at energies above 10 MeV, which differ significantly from SRIM calculations. Simultaneously, the approximations \cite{Barbui10par,Knyazheva06} give us significantly smaller $SP$ values. As may also result from the figure, a lack of data for very heavy ions at low energies has led to very small $SP$ values at energies below 10 MeV, where the nuclear (collisional) component has appeared according to the approximations \cite{SRIM,Ziegler80}. This component might be significantly smaller than it follows from the approximations, as discovered in the analysis of carbon stopping-power data at low energies \cite{Saga23}. Note that the nuclear component plays a key role at the end of a HI range since it determines the final values of a mean range and the respective range struggling, and thus cannot be neglected. TRIM simulations (Monte Carlo version of SRIM) that take into account evaporation of particles from a compound nucleus (CN) and ERs stopping inside a target \cite{Saga13} might be the starting point in solving the problem.

\section{TRIM simulations of ER ranges in Mylar}
\label{TRIMsim}

\begin{figure}[h]  
\centering
\includegraphics[width=0.85\textwidth]{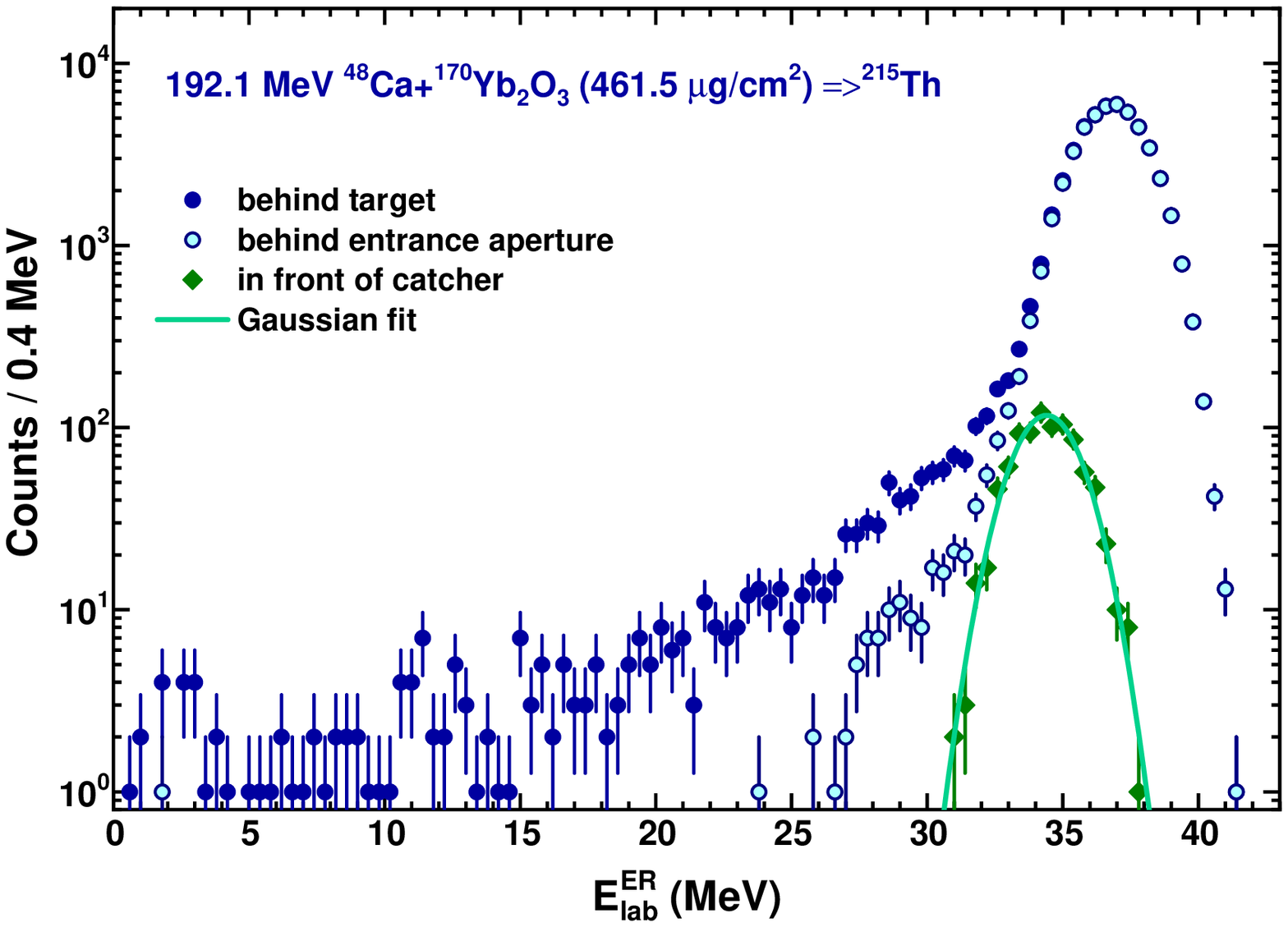}
\vspace{-2.0mm}
\caption{The results of TRIM simulations are shown for the output energies of $^{215}$Th escaping the target (solid circles), passing through the entrance aperture of GASSOL (open circles), slowing down in hydrogen, and cutting by a catcher radius (solid diamonds), i.e., just in front of the catcher. The Gaussian fit to the last distribution is shown by a solid line.}\label{EsimTRIM}
\end{figure}

The energy distribution of $^{215}$Th ERs produced in the $^{48}$Ca + $^{170}$Yb reaction at the input energy of $^{48}$Ca 192.1 MeV in front of the 461.5 $\mu$g/cm$^{2}$ $^{170}$Yb$_{2}$O$_{3}$ target was simulated with TRIM, considering evaporation of 3 neutrons from the  $^{218}$Th$^{*}$ CN and $^{215}$Th stopping inside the target layer. The average neutron energies, the position of the input $^{48}$Ca energy relative to the 3n-evaporation excitation function calculated with HIVAP \cite{HIVAP}, and the projectile stopping inside the target layer were taken into account. Fig.~\ref{EsimTRIM} shows the results of 5$\cdot$10$^{4}$ simulations for $^{215}$Th births inside the target, which led to output energies for atoms escaping the target, passing through the entrance aperture of GASSOL, and slowing down in hydrogen, i.e., the energies in front of the Mylar catcher. The small yield for the last values corresponds to the cut of events by the catcher of a defined radius, since the focusing of charged ERs by the magnetic field of GASSOL was not taken into account. Note that variation of the catcher radius in a reasonable range did not change the values of the average energy and width within their errors, as obtained by the Gaussian fitting of these distributions.

\begin{figure}[h]  
\centering
\includegraphics[width=0.85\textwidth]{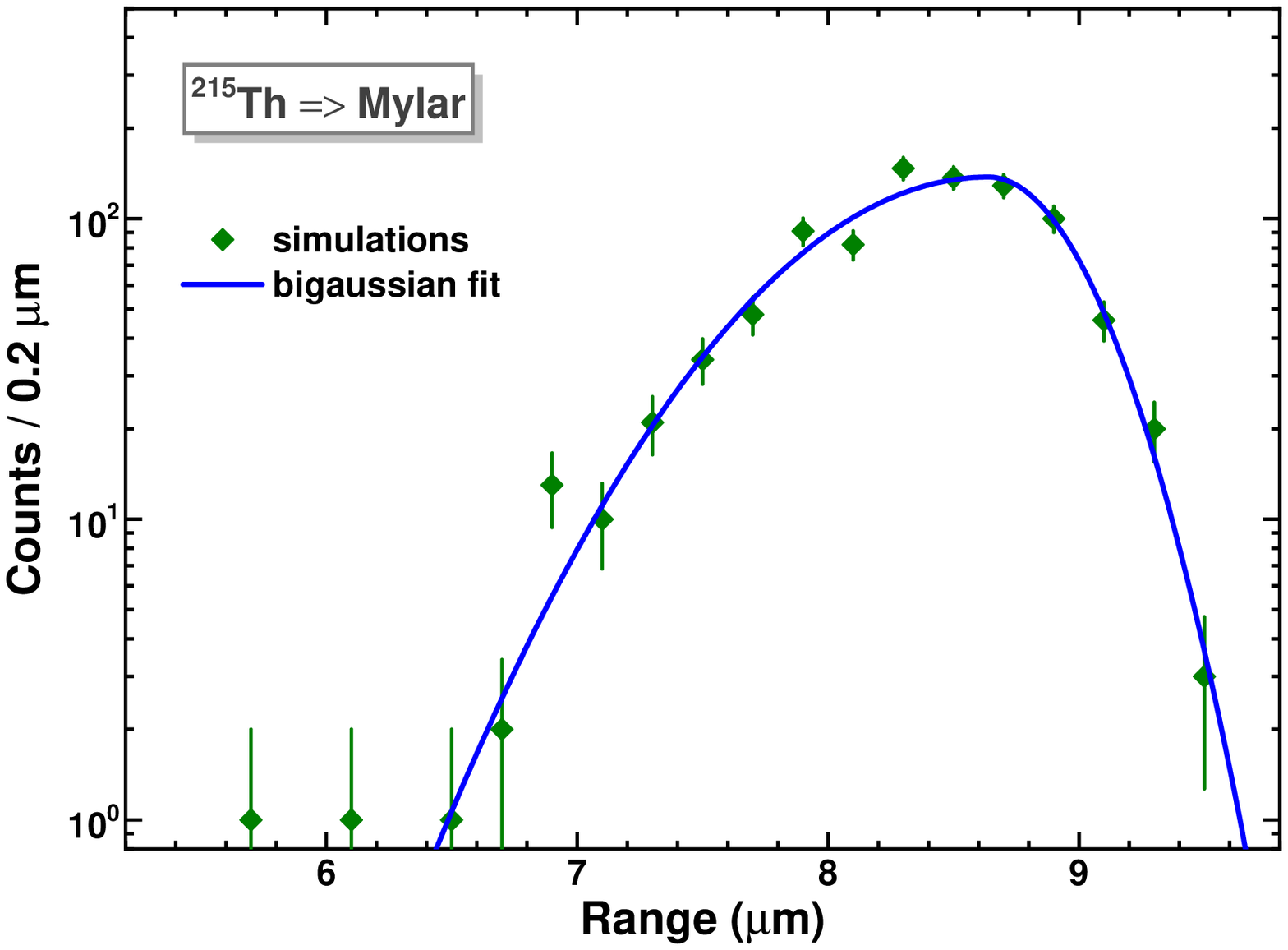}
\vspace{-2.0mm}
\caption{The range distribution of $^{215}$Th atoms, having the energies cut by a catcher's radius, as depicted in Fig.~\ref{EsimTRIM}, is shown by solid diamonds. The result of the 4-parameter bigaussian fit, corresponding to Eq.~(\ref{bigauss}), is shown by a solid line.}\label{TRIMrange}
\end{figure}

In Fig.~\ref{TRIMrange}, the results of TRIM simulations are shown for the range distribution of $^{215}$Th atoms having the cut energies depicted in Fig.~\ref{EsimTRIM} and a forward direction towards the beam \cite{Saga13}. The obtained range distribution was approximated by the asymmetric 4-parameter bigaussian function, having two different widths $w_{1}$ and $w_{2}$, and  the same values for amplitude $H$ and average range $R_{m}$ as the fitted parameters:
\begin{eqnarray}\label{bigauss}
   R & = & H \exp\{-[(R - R_{m}) / (2 w_{1})]^2\},  \mbox{if } R <  R_{m}\\
\nonumber
       & = & H \exp\{-[(R - R_{m}) / (2 w_{2})]^2\}, \mbox{if } R \geqslant  R_{m}.
\end{eqnarray}
This approximation will be useful further for the estimation of range distributions using different stopping power approaches considered above. allowing us to derive the value of the average range only.

\section{ER ranges derived from stopping power approximations}
\label{rangesSP}

The derivation of ranges from stopping powers was performed using SRIM calculations (Section~\ref{SRIMappr}), Ziegler's approximations (Section~\ref{Zieglerappr}), and 2p- and 4p-approximations obtained in the present work (Sections~\ref{2parappr} and \ref{4parappr}, respectively). These approaches describe low-energy $SP$s better than others (see Sections~\ref{5parappr} and \ref{Bohrappr}), even though $SP$ values from the last two approximations (2p and 4p) seem to be underestimated since they appear to be insensible to nuclear (collisional) stopping. In the average range estimates below, the nuclear stopping component calculated by SRIM was added to the approximated $SP$ values. This component could be overestimated, as was discovered for carbon $SP$ in \cite{Saga23} and mentioned above.

To check the correctness of the numerical integration of the stopping power curves shown in Fig.~\ref{215ThSP}, total $SP$ values calculated by SRIM were integrated for the input energy of $^{215}$Th $E^{\rm ER}_{m} = 34.41$ MeV derived from the Gaussian fitting the energy distribution obtained in TRIM simulations (see Fig.~\ref{EsimTRIM}). The obtained integrated value according to Eq.~(\ref{range}), $R^{\rm SRIM}_{mi} = 1.171$ mg/cm$^{2}$, was close to the value given by SRIM $R^{\rm SRIM}_{p} = 1.175$ mg/cm$^{2}$. The difference between the values was attributed to inaccuracy in numerical integration. Note that SRIM gives an average value of the HI projected range, $R^{\rm SRIM}_{p}$, whereas the $R^{\rm SRIM}_{mi}$ value might be considered as the value of a total (traveling) range obtained without knowledge of HI  scattering inside a matter. Despite this, a small $A_{m}/A_{\rm HI}$ value, allows us to neglect the difference between $R_{p}$ and $R_{mi}$ in further consideration, guided by the theory \cite{LSS63}.

\begin{figure}[h]  
\centering
\includegraphics[width=0.85\textwidth]{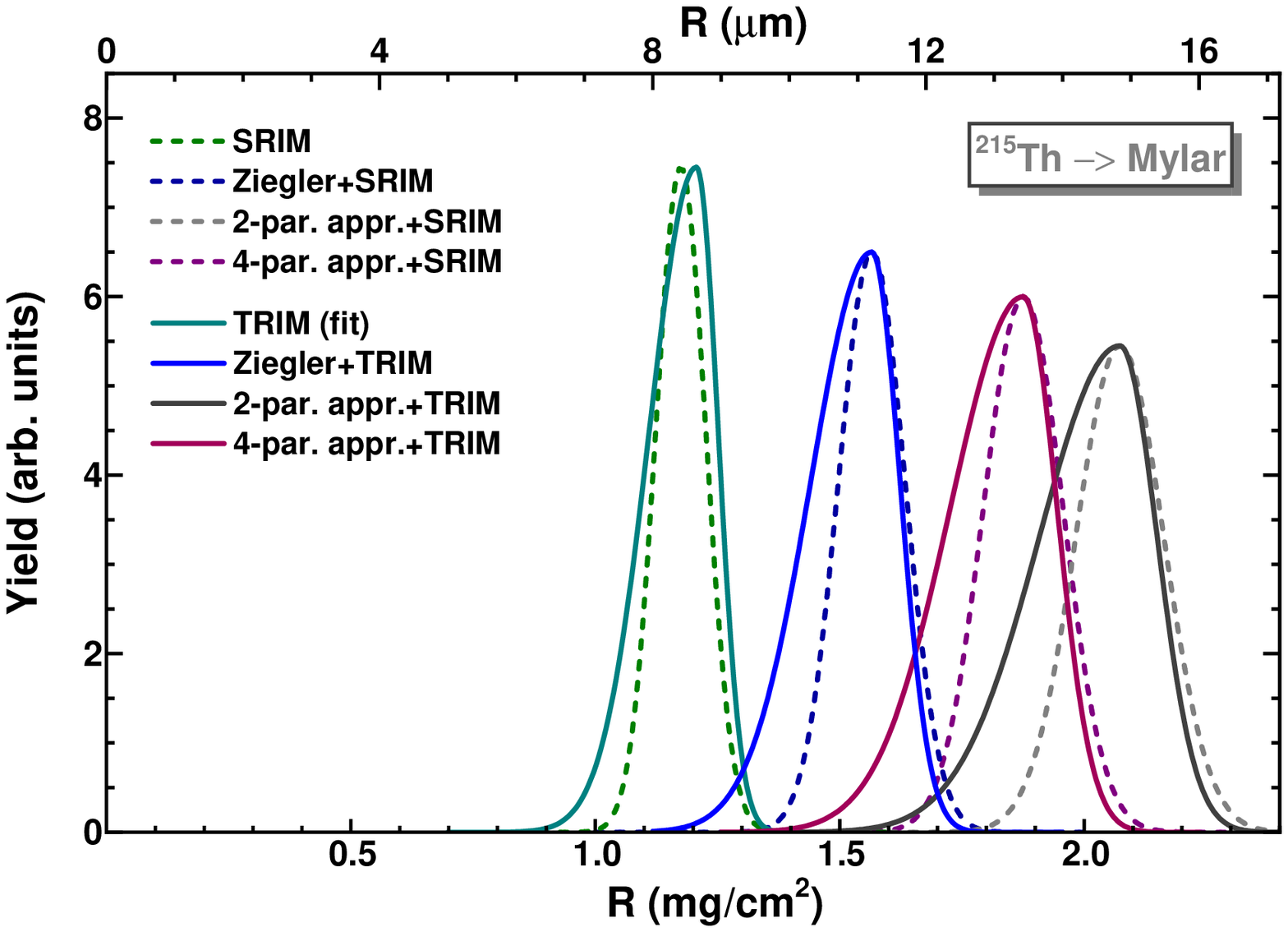}
\vspace{-2.0mm}
\caption{The range distribution of $^{215}$Th atoms, as obtained by integrating stopping powers calculated in different approximations (see Fig.~\ref{215ThSP}). Dashed curves of different colors correspond to Gaussian fits with average range values $R_{mi}$ and widths proportional to these values and the relative width given by SRIM for the $^{215}$Th average energy $E^{\rm ER}_{m}$. Solid curves of similar colors correspond to the correction of these widths introduced by accounting for the evaporation of neutrons and stopping of $^{215}$Th ERs inside the target layer. This  leads to the energy distribution shown in Fig.~\ref{EsimTRIM} and the respective range distribution shown in Fig.~\ref{TRIMrange}. See more details in the text.}\label{Thranges}
\end{figure}

Fig.~\ref{Thranges} shows the range distributions of $^{215}$Th, which were derived by integrating stopping powers calculated in different approximations (see Fig.~\ref{215ThSP}). As a result, the SRIM distribution had the smallest amounts of the average range $R_{m}^{\rm SRIM}$ and width $ w_{\rm SRIM}$ calculated for the mean energy $E^{\rm ER}_{m} = 34.41$ MeV (see Fig.~\ref{EsimTRIM}). This distribution is compared to the one obtained in TRIM simulations and fitted by the bigaussian (see Fig.~\ref{TRIMrange}). It maybe reminded that the simulated energy distribution accounting for the evaporation of neutrons and stopping of $^{215}$Th ERs inside the target layer (see Fig.~\ref{EsimTRIM}) was used in these simulations. Larger ranges were derived for integrated stopping power approximations. As mentioned above, integrating gives the average range value, $R_{mi}$. For the widths, simple scaling  was used to account for the evaporation of neutrons and stopping of $^{215}$Th ERs inside the target layer, as obtained in TRIM simulations. With the results of bigaussian fitting using Eq.~(\ref{bigauss}) these widths are expressed as $w_{1} = w^{\rm TRIM}_{1} R_{mi} / R_{m}^{\rm TRIM}$ and $w_{2} = w^{\rm TRIM}_{2} R_{mi} / R_{m}^{\rm TRIM}$.

As we see in Fig.~\ref{Thranges}, different approaches results in a wide variation in average values and widths of distributions that covers a range from 6.5 to 16.5 $\mu$m, in which $^{215}$Th ERs may completely stop in Mylar. This difference in range estimates implies the use of catcher foils of a respective thickness to collect certainly all ERs produced in the reaction. Any catcher thickness deteriorate the resolution of $\alpha$-peaks in the detected $\alpha$-spectrum from the collected ERs. Present conditions seem worse than those that took place for the detection of $\alpha$-particles from Rn ERs produced in the $^{16}$O+$^{194}$Pt reaction, which were collected by the Al rotating disc \cite{Saga19}. Note that in this case, the Rn ranges in Al at energies of 5.5--9 MeV exceeded those calculated by SRIM by a factor of 1.65.

\section{Simulations of $\alpha$-spectra detected for different ER ranges}
\label{alphasim}

In thought experiments, $^{215}$Th was chosen as the best detectable product produced in the $^{48}$Ca+$^{170}$Yb reaction due to its suitable half-life (1.2 s) and $\alpha$-lines of 7.334 (8\%), 7.392 (52\%) and 7.522 (40\%) MeV \cite{215Th}. This nuclide was well detected in test experiments carried out with DGFRS-2 \cite{DGFRS2} using the same reaction. In experiments with the rotating catcher assumed for testing GASSOL, the detection of $\alpha$-particles is complicated by uncertainty in the range of $^{215}$Th, as it follows from Fig.~\ref{Thranges}. Using Monte Carlo simulations, an attempt to evaluate degradation of the $^{215}$Th $\alpha$-energies was performed with the fixed values of the catcher central radius (75 mm) and sizes of the semiconductor detector (29$\times$34 mm). As for ranges, extreme cases were examined under the assumption of the TRIM range distribution and catcher thicknesses of $w_{c}$ = 11 and 17 $\mu$m. The first value was determined by the TRIM range distribution, whereas the second one corresponded to the range distribution derived from the 2-parameter $SP$ approximation (see Fig.~\ref{Thranges}). It could be expected that $\alpha$-energies obtained in this case would be similar to those emerging from the 11-$\mu$m catcher foil that collected $^{215}$Th ERs having the TRIM range distribution. Four additional options corresponding to the ``good'' and ``bad'' focusing of ERs (the track width for the collected atoms $b_{c}$ = 20 and 50 mm, respectively) were examined. Two different distances between the catcher and detector, corresponding to $L_{cd}$ = 10 and 40 mm, were also considered.

\begin{figure}[ht!]  
\begin{center}
\includegraphics[width=0.85\textwidth]{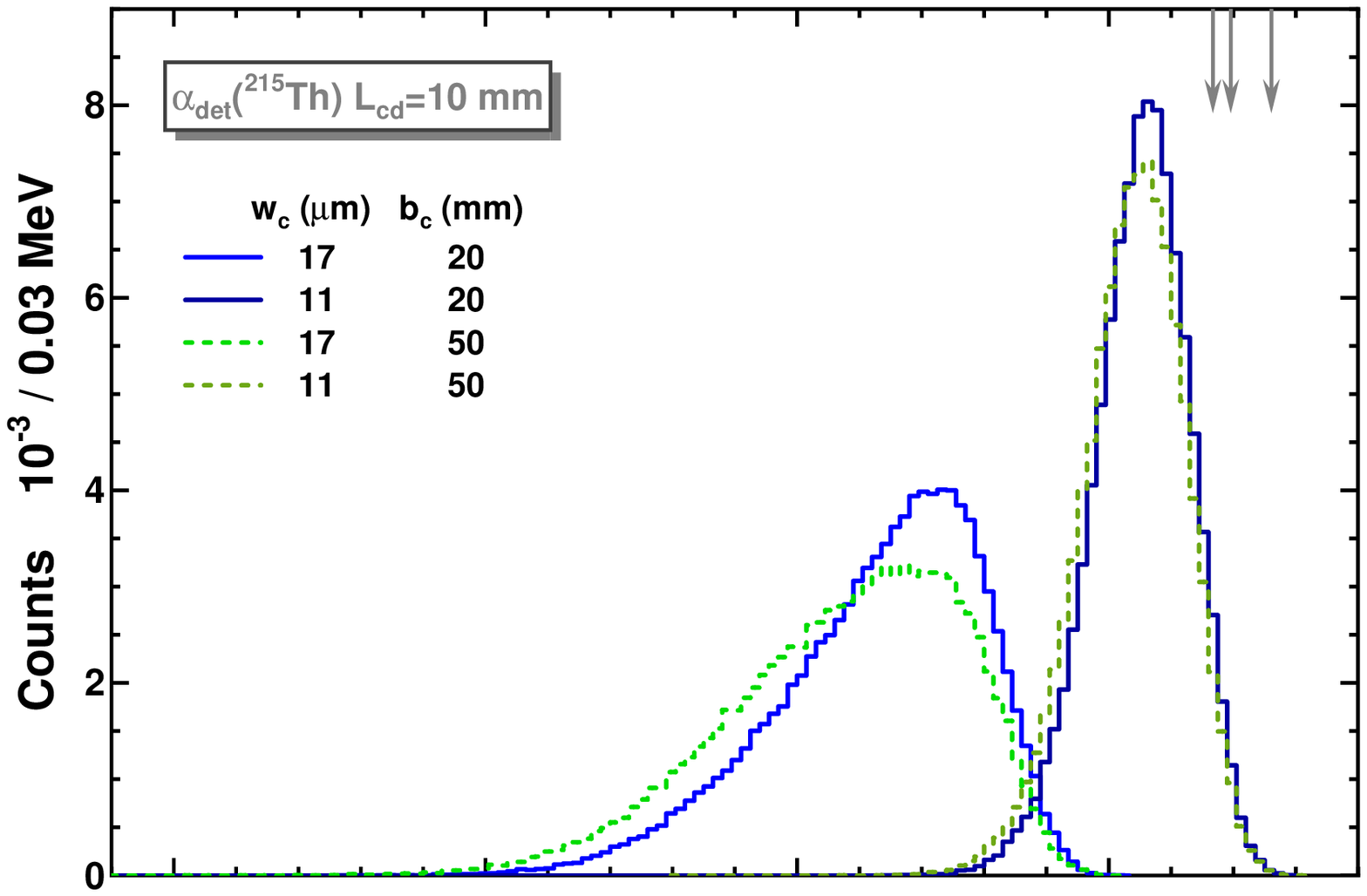}\vspace{-11.5mm}
\includegraphics[width=0.85\textwidth]{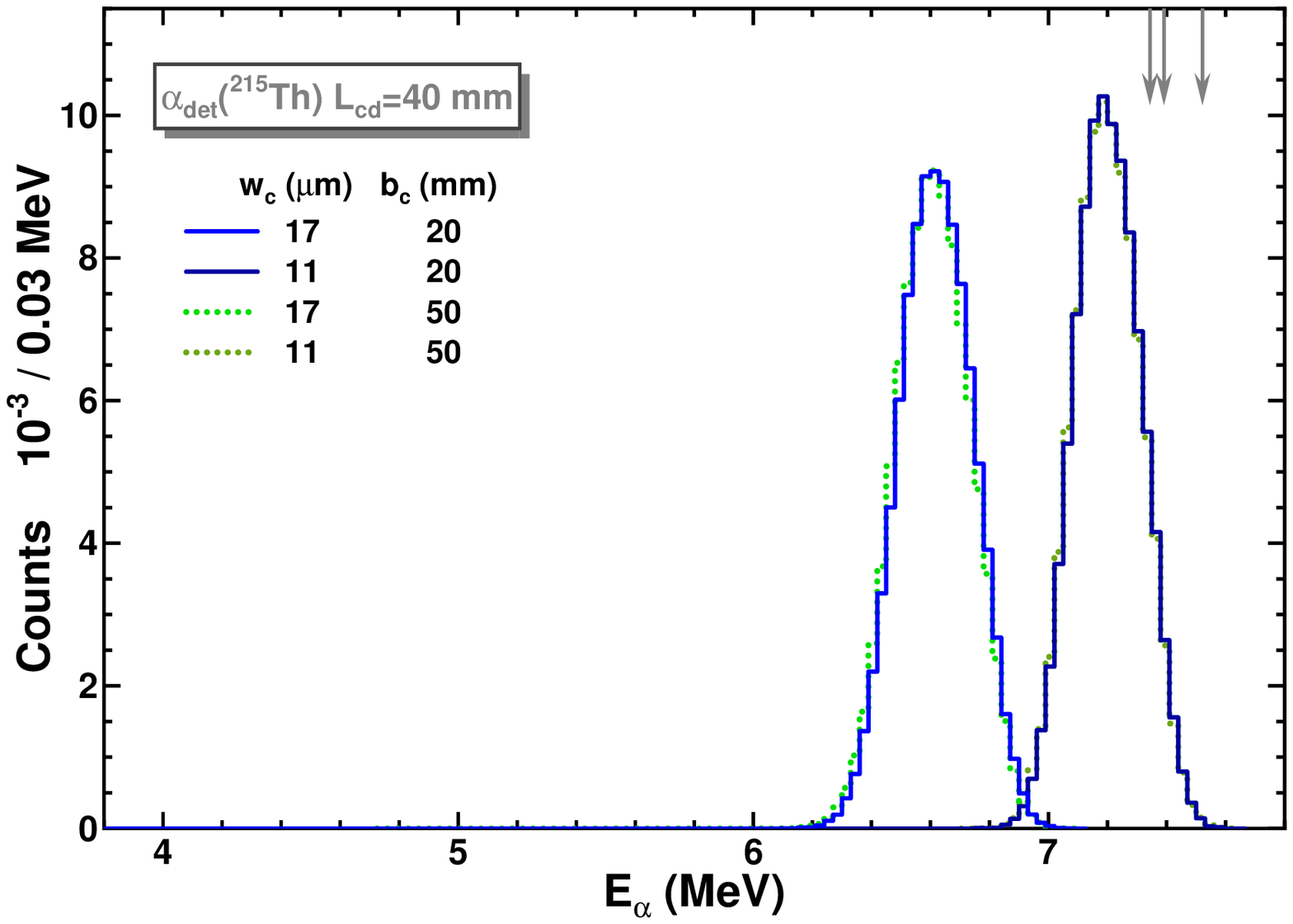}
\end{center}
\vspace{-9.0mm}
\caption{The detected $\alpha$-energies of $^{215}$Th ERs, as obtained with TRIM simulations, are shown. Spectra correspond to the TRIM range distribution shown in Fig.~\ref{TRIMrange} and different  detection conditions for emerged $\alpha$-particles (indicated in the figure). Arrows mean the original  $^{215}$Th $\alpha$-energies. See details in the text.}\label{EaLcd1040}
\end{figure}

Fig.~\ref{EaLcd1040} shows the results of 10$^{5}$ TRIM simulations for $^{215}$Th $\alpha$-particles detected by the semiconductor detector with a native resolution of $\sigma_{E}$ = 50 keV. Different spectra thus obtained correspond to the TRIM range distribution shown in Fig.~\ref{TRIMrange} and different conditions for the detection of emerged $\alpha$-particles, which are as follows:  $L_{cd}$ = 10 mm, $w_{c}$ = 11 and 17 $\mu$m, $b_{c}$ = 20 and 50 mm (upper panel), and $L_{cd}$ = 40 mm with the same $w_{c}$ and $b_{c}$ (bottom panel). As one can see in the figure, and not surprisingly, total widths for the resulting spectra for the remote catcher are less than those for the close one, although $\alpha$-lines of  $^{215}$Th are not resolved in both cases. Significant energy losses take place in the case of a thick catcher ($w_{c}$ = 17 $\mu$m) and a wide track width ($b_{c}$ = 50 mm). Note that the detection of $^{215}$Th $\alpha$-particles should not interfere with the alphas of neighboring isotopes $^{214}$Th and $^{216}$Th that also could be produced in respective neutron evaporation reactions. Their yields could be suppressed by a proper choice of the rotation velocity of the catcher, bearing in mind the shorter half-lives for these isotopes (87 and 26 ms, respectively).

\section{Summary}
\label{summa}

A lack of data on ranges of heavy ions (HIs) in Mylar, which could be used for practical purposes, prompted the use of available stopping power ($SP$) data for HIs \cite{IAEAsp}, implying their subsequent approximation and integration over energy for range estimates. The comparison of these data with the available semi-empirical approaches \cite{SRIM,Saga15,Ziegler80,Barbui10par,Knyazheva06}, including the one proposed in the present work, showed unsatisfactory reproduction of some $SP$ data at low HI energies. Applying the semi-empirical approximations to the $SP$ estimates of very heavy evaporation residues (ERs) produced in complete fusion reactions with HIs, such as $^{215}$Th, it seems that the nuclear (collisional) $SP_{n}$ was not taken into account in the most recent ones \cite{Saga15,Barbui10par,Knyazheva06}. One more observation, to which attention can be paid, is larger values of the electronic stopping power $SP_{e}$ predicted by SRIM \cite{SRIM} in comparison with other approximations at energies below 10 MeV. Along with that, differences in the total $SP$ values obtained in the considered approaches could also be explained by a lack of low-energy data for a proper approximation. Another explanation could be smaller values of nuclear stopping than it is used in the earlier approaches \cite{SRIM,Ziegler80}.

The integration of the inverse values of the total $SP$ calculated according to \cite{SRIM,Saga15,Ziegler80} and obtained in the present work for $^{215}$Th, as an example of an evaporation residue (ER) produced in fusion reactions, was performed. The initial energy corresponding to its average value derived from TRIM simulations was used. Energy losses in the target layer and hydrogen gas were taken into account. Four different range distributions were obtained as a result. The largest value of the average range determined with the 2-parameter approximation \cite{Saga15} differs from the smallest one according to SRIM by a factor of 1.7. Given these differences, there may be difficulties in choosing the proper thicknesses of the pre-stopper and catcher foils, implying a total collection of ERs and providing minimal energy losses of $\alpha$-particles inside the catcher to identify ERs. TRIM simulations for the $^{215}$Th $\alpha$-particles emitted from the catcher showed that the distance between the catcher and detector is a crucial value, ensuring the smallest width of $\alpha$-lines in detected spectra if the catcher is a remote distance from the detector.

In conclusion, it may also be noted that using a thick catcher and varying the range distribution in simulations, one can determine true ranges of $^{215}$Th ERs by comparing measured $\alpha$-spectra with the simulated ones, as was done in \cite{Saga19}. These ranges can also be derived by measuring an integral range distribution through the variation of the thickness of a pre-stopper (degrader) foil installed  in front of the catcher position, with subsequent differentiation of this distribution. The comparison of both values is of particular interest.

\section{Acknowledgements}

The author wishes to acknowledge Dr. D.I. Solovyev for his interest in the work and useful remarks and comments.

\end{document}